\documentclass[aps,pre,twocolumn,groupedaddress,longbibliography,nofootinbib,showkeys,showpacs]{revtex4-2}
\usepackage{latexsym}
\usepackage{amssymb}
\usepackage{amsthm}
\usepackage{amsmath}
\usepackage{amsfonts}
\usepackage{graphicx}
\usepackage{mathptmx}      % use Times fonts if available on your TeX system
\usepackage{bm}% bold math symbols \bm\alpha
\usepackage{enumerate}
\usepackage{subfigure}
\usepackage{color}
\usepackage{hyperref}
\usepackage{listings}
\usepackage{grffile}
\definecolor{DarkGreen}{RGB}{0,100,0}
\begin{document}

%\preprint{}

\title{On the hardness of quadratic unconstrained binary optimization problems}

\author{V. Mehta\footnote{ORCID: 0000-0002-9123-7497}}
\affiliation{Institute for Advanced Simulation, J\"ulich Supercomputing Centre,\\
Forschungszentrum J\"ulich, D-52425 J\"ulich, Germany}
\author{F. Jin\footnote{ORCID: 0000-0003-3476-524X}}
\affiliation{Institute for Advanced Simulation, J\"ulich Supercomputing Centre,\\
Forschungszentrum J\"ulich, D-52425 J\"ulich, Germany}
\author{K. Michielsen\footnote{ORCID: 0000-0003-1444-4262}}
\affiliation{Institute for Advanced Simulation, J\"ulich Supercomputing Centre,\\
Forschungszentrum J\"ulich, D-52425 J\"ulich, Germany,}
\affiliation{AIDAS, 52425 J\"ulich, Germany,}
\affiliation{RWTH Aachen University, 52056 Aachen, Germany}
\author{H. De Raedt\footnote{ORCID: 0000-0001-8461-4015}}
\affiliation{Institute for Advanced Simulation, J\"ulich Supercomputing Centre,\\
Forschungszentrum J\"ulich, D-52425 J\"ulich, Germany}
\affiliation{Zernike Institute for Advanced Materials,\\
University of Groningen, Nijenborgh 4, NL-9747 AG Groningen, Netherlands}

\date{\today}

\begin{abstract}
We use exact enumeration to
characterize the solutions of quadratic unconstrained binary optimization problems
of less than 21 variables in terms of their distributions of Hamming distances to close-by solutions.
We also perform experiments with the D-Wave Advantage 5.1 quantum annealer, solving many instances
of up to 170-variable, quadratic unconstrained binary optimization problems.
Our results demonstrate that the exponents
characterizing the success probability of a D-Wave annealer to solve a QUBO
correlate very well with the predictions based on the Hamming distance distributions
computed for small problem instances.
\end{abstract}

%\begin{keyword} %% keywords here, in the form: keyword \sep keyword
%quantum computation\sep computer simulation\sep high performance computing\sep parallelization\sep  benchmarking
%\PACS{03.67.Lx, 02.70.-c}
%\end{keyword}
\date{\today}

\maketitle
%\end{frontmatter}
%\tableofcontents

\section{Introduction}

Optimization is at the heart of problem solving in science, engineering, finance, operational research etc.
The basic idea is to associate a cost with each of the possible
values of the variables that describe the problem and try to minimize this cost.
Of particular importance is the class of so-called discrete optimization problems in
which some or all the variables take values from a finite set of possibilities.
Discrete optimization problems are often NP-hard~\cite{GARE00} which, in practice and in simple terms, means that
solving such a problem on a digital computer will require resources
that increase exponentially with the number of variables.

Many discrete optimization problems can be reformulated
as quadratic unconstrained binary optimization (QUBO) problems~\cite{LUCA14,Lewis2017}.
Solving a QUBO amounts to finding the values of the $N$ binary variables $x_i=0,1$ that minimize the cost function
\begin{eqnarray}
\hbox{Cost}(x_1,\ldots,x_{N})&=&\sum_{1=i\le j=N} Q_{i,j}x_i \,x_j \quad,\quad x_i=0,1
\;,
\label{QUBO0}
\end{eqnarray}
where $Q_{i,j}=Q_{j,i}$ is a symmetric $N\times N$ matrix of floating point numbers.

The interest in expressing discrete optimization problems in the form of QUBOs has recently gained momentum
by the development of quantum annealers manufactured by D-Wave systems~\cite{johnson2011quantum,mcgeochd}.
In theory~\cite{Kadowaki1998}, quantum annealers make use of the adiabatic theorem~\cite{Born1928}
to find the ground state of the Ising model defined by the Hamiltonian
\begin{eqnarray}
H&=&\sum_{1=i<j=N} J_{i,j}S_i\,S_j +  \sum_{i=1}^N h_i S_i
\;,
\label{QUBO1}
\end{eqnarray}
Substituting $S_i=1-2x_i=\pm1$ in Eq.~(\ref{QUBO1}) yields
\begin{eqnarray}
H=\hbox{Cost}(x_1,\ldots,x_{N})-C_0
\;,
\label{QUBO1a}
\end{eqnarray}
where the relations between the $Q$'s in Eq.~(\ref{QUBO0}) and
$J$'s, $h's$, and $C_0$ in Eqs.~(\ref{QUBO1}) and~(\ref{QUBO1a}) are given in appendix~\ref{appB}.

Obviously, the transformation $S_i=1-2x_i=\pm1$ does not change the nature of the optimization problem.
In other words, minimizing the cost function of a QUBO Eq.~(\ref{QUBO0}) or
finding the ground state of the Ising model Eq.~(\ref{QUBO1}) are equally difficult.
Currently available quantum annealer hardware often finds ground states of Eq.~(\ref{QUBO1})
for fully-connected problems with about 200 variables or less in about micro seconds~\cite{johnson2011quantum,mcgeochd},
which is quite fast, suggesting that as larger quantum annealers become
available, they have the potential to solve large QUBOs in a relatively short real time.

Although the transformation $S_i=1-2x_i=\pm1$ does not change the nature of the optimization problem,
the formulation of a particular optimization problem in terms of a QUBO can.
For instance, 2-satisfiability problems (2SAT)~\cite{GARE00} (see section~\ref{SAT})
can be solved with computational resources that increase linearly
with the number of binary variables~\cite{Krom1967,Even1976,Aspvall1979}.
However, the special features of 2SAT-problem that permits its efficient solution are lost
when it is expressed as a QUBO/Ising model.
In fact, the equivalent QUBO becomes notoriously hard to solve by e.g., simulated annealing~\cite{neuhaus2014monte}.
On the other hand, it is also not difficult to construct Ising models of which the ground state
is very easy to find.

In view of the potential of quantum annealers for solving large QUBOs in the near future,
it is of interest to gain some insight into the
degree of success by which a quantum annealer is expected to
solve a QUBO/finding the ground state of corresponding Ising model,
without actually performing the experiment.

In this paper, we show that there are at least three different classes of QUBOs
that distinguish themselves by
\begin{enumerate}
\item
The success probability a D-Wave annealer finds the ground state of the Ising model/solves the QUBO.
\item
The distribution of Hamming distances between the ground state and the lowest excited states
computed for relatively small, representative problem instances.
\end{enumerate}
We demonstrate that the differences between the Hamming distance distributions are correlated with
the size-dependent scaling of the success probabilities with which D-Wave quantum annealers
find the ground state.

The structure of the paper is as follows.
Section~\ref{PROB} introduces the three different classes of QUBO problems that we analyze.
In section~\ref{METH}, we briefly review the three different methods by which we solve the QUBO instances.
Sections~\ref{RESU} and~\ref{RQA} present and discuss our results for the Hamming distance and level
spacing distributions, and quantum annealing experiments, respectively.
In section~\ref{CONC}, we summarize our findings.

\section{Quadratic Unconstrained Binary Optimization problems}\label{PROB}

\subsection{2-Satisfiability problems}\label{SAT}

The problem of assigning values to binary variables
such that given constraints on pairs of variables are satisfied is called 2-satisfiability~\cite{GARE00}.
2SAT is a special case of the general Boolean satisfiability problem,
involving constraints on more than two variables.
In contrast to e.g. 3SAT which is known to be NP-complete,
2SAT can be solved in polynomial time.
The most efficient algorithms solve 2SAT in a time which is proportional to the
number of variables~\cite{Krom1967,Even1976,Aspvall1979}.

A 2SAT problem is specified by $N$ binary variables $x_i=0,1$ and a conjunction of $M$ clauses
defining a binary-valued cost function
\begin{eqnarray}
C &=& C(x_1,\ldots,x_N)
\nonumber \\
&=& (L_{1,1} \lor L_{1,2}) \land (L_{2,1} \lor L_{2,2}) \land ... \land (L_{M,1}
\lor L_{M,2})
\;,
\label{H2SAT0}
\end{eqnarray}
where the literal $L_{\alpha,j}$ stands for either
$x_{i(\alpha,j)}$ or its negation $\overline{x}_{i(\alpha,j)}$, for $\alpha=1,...,M$ and $j=1,2$.
The function $i(\alpha,j)$ maps the pair of indices $(\alpha,j)$ onto the index $i$ of the binary variable $x_i$.
A 2SAT problem is satisfiable if one can find at least one assignment of the $x_i$'s which
makes the cost function $C$ true.

Solving a 2SAT problem is equivalent to finding the ground state of the Ising-spin
Hamiltonian~\cite{neuhaus2014quantum,neuhaus2014monte,LUCA14}
\begin{equation}
H_{\mathrm{2SAT}} =
%H_{\mathrm{2SAT}}(S_1,\ldots,S_N) =
\sum_{\alpha=1}^M h_{\mathrm{2SAT}}(\epsilon_{\alpha,1}S_{i(\alpha,1)},\epsilon_{\alpha,2}S_{i(\alpha,2)})
\;,
\label{H2SAT1}
\end{equation}
where $\epsilon_{\alpha,j}=+1(-1)$ if $L_{\alpha,j}$ stands for $x_i\,(\overline{x}_i)$
and
\begin{equation}
h_{\mathrm{2SAT}}(S_l,S_m)=(S_l-1)(S_m-1)\quad,\quad S_m,S_l=\pm1
\;.
\label{H2SAT2a}
\end{equation}
Grouping and rearranging terms, Eq.~(\ref{H2SAT1}) reads
\begin{equation}
H_{\mathrm{2SAT}} = \sum_{1\le i<j\le N} J_{i,j}S_i S_j + \sum_{i=1}^{N} h_i S_i +C_1
\;,
\label{H2SAT2}
\end{equation}
where $C_1$ is an irrelevant constant.
Therefore, solving a 2SAT problem Eq.~(\ref{H2SAT0}) is equivalent
to solving the QUBO problem defined by Eq.~(\ref{H2SAT2}).
It may be of interest to mention that if one is given a QUBO problem without
knowing that it originated from a 2SAT problem, it is not clear that
the QUBO can be solved in a time linear in $N$.

Constructing 2SAT problems is easy but finding 2SAT problems that have a unique known ground state
and a highly degenerate first-excited state quickly becomes more difficult with increasing $N$~\cite{neuhaus2014monte}.
We have generated 15 sets of 2SAT problems that have a unique known ground state
and a highly degenerate first-excited state, each set corresponding to an $N$,
with $N$ ranging from 6 to 20 with the number of clauses $M=N+1$.
The graphs representing these problems are not fully connected, that is not all $J_{i,j}$'s
are different from zero.
For our sets of 2SAT problems, the $h_{i}$'s and $J_{i,j}$'s can take the integer values between $-2$ and $2$.

\subsection{Fully-connected spin glass}\label{SPIN}

The spin glass model is defined by the Hamiltonian  Eq.~(\ref{QUBO1}).
Computing the ground state configuration is, in general, very hard.
To verify that a spin configuration has the lowest energy,
one would (in general) have to go through all the $2^N-1$ other configurations to check if
it indeed has the lowest energy.
The qualifier ``in general'' is important here for there are cases, such when all $J_{i,j}=0$,
for which the ground state is trivial to find.
To effectively rule out such trivially solvable problems,
we use uniform (pseudo) random numbers in the range $[-1,+1]$
to assign values to all the $J_{i,j}$'s and all the $h_{i}$'s.
The probability that one of the $J_{i,j}$'s is zero is extremely small,
justifying the term ``fully-connected spin glass''.
In the following, we refer to the set of model instances generated in this manner as RAN problems.

\subsection{Fully-connected regular spin-glass model}\label{REGU}

In the course of developing an QUBO-based application to benchmark large clusters of GPUs (see appendix~\ref{appA}),
we discovered by accident that the ground state of the spin glass defined by Eq.~(\ref{QUBO1}) with
\begin{eqnarray}
J_{i,j} &=& 1- (i+j-2)/(N-1)\quad,\quad i\not=j
\;,
\nonumber \\
h_i&=&1-2(i-1)/(N-1)
\;,
\label{SPIN0a}
\end{eqnarray}
seems to have a peculiar structure.
Note that of the order of $N$ $J_{i,j}$'s are zero.

Although we have not been able to give a proof valid for all $N$, up to $N=200$
we have not found any counter example for conjecture
that the ground states of the Ising model with parameters given by Eq.~(\ref{SPIN0a})
is given by $(S_1=-1,...,S_k=-1,S_{k+1}=1,\ldots,S_N=1)$
or, equivalently $(x_1=1,...,x_k=1,x_{k+1}=0,\ldots,x_N=0)$
where $k$ is the integer that minimizes
\begin{eqnarray}
f(k)=\sum_{1\le i \le j\le k}Q_{i,j}= \frac{k (N-k) (N-2k+2)}{N-1}
\;.
\label{SPIN1}
\end{eqnarray}
Thus, although we are solving a fully connected QUBO problem, if our conjecture is correct,
its solution is very easy to find for any $N$.
We refer to the special, fully-connected regular spin glass problems defined
by Eq.~(\ref{SPIN0a}) as REG problems

From Eq.~(\ref{QUBO1}), it immediately follows that randomly reversing a spin $i$ and replacing $h_j$ by $-h_j$ and $J_{i,j}$ by $-J_{i,j}$
for all $j$ does not change the ground state energy of a QUBO problem.
Applying such ``gauge transformation'' to a sets of randomly selected spins
generated a set of REG problems that are mathematically equivalent.
This feature, in combination with the fact that the ground state is known (for at least $N\le200$) makes
REG problems well-suited for testing and benchmarking purposes, of both conventional and quantum hardware.

\section{Methods for solving QUBOs}\label{METH}

We solve QUBOs using three different methods.

\begin{enumerate}
\item
A computer code referred to as {\bf QUBO22} which uses GPUs and/or CPUs
to solve QUBOs, Polynomial Unconstrained Binary Optimization (PUBOs),
and Exact Cover problems (reformulated as QUBOs).
{\bf QUBO22} simply enumerates all $2^N$ possible values of the binary variables $x_1,\ldots,x_N$
while keeping track of those configurations of $x$'s that yield the lowest, next to lowest and largest cost.
{\bf QUBO22} obviously always finds the true ground state.
The number of arithmetic operations required to solve a QUBO is proportional to $N(N-1)2^N$.
With the supercomputers that are available to us,
the exponential increase with $N$ limits the application {\bf QUBO22} to problems of size $N\le56$.

In order to compute the Hamming distance between excited states and the ground state
and level spacing distributions, it is necessary to keep track of a large number of different states.
To this end, we use another code, also based on full enumeration, which in practice, can readily
handle problems up to $N=20$.

\item
Heuristic methods can solve QUBOs in (much) less time than {\bf QUBO22} can.
However, heuristic methods do not guarantee to return the solution of the QUBO problem (although
they very often do).
In our work, we use {\bf qbsolv}, a heuristic solver provided by D-Wave, to compute the ground states
of all problem instances.
For those problems which {\bf QUBO22} can solve, the ground states obtained by {\bf qbsolv} and {\bf QUBO22} match.
For all RAN and REG problems up to $N=200$,
the ground states obtained by {\bf qbsolv} and the D-Wave Advantage 4.1 Hybrid solver are also the same.

\item
We have used the D-Wave Advantage 5.1 quantum annealer to solve all problem instances.
We calculate the success probabilities by using the ground states obtained by {\bf qbsolv}.
The data for the success probabilities is then used to analyse the scaling behavior as a function
of the problem size $N$.

\end{enumerate}

\section{Hamming distance and level spacing}\label{HAMMI}\label{RESU}

The Hamming distance between two bitstrings (or strings of $S$'s) of equal length is defined as
the number of positions at which the corresponding bits ($S$'s) are different.
For each of the problems in our set, we use exact enumeration to find at most 6037 states with the lowest energies.
With this data we compute the Hamming distances between the ground state and these excited states.

In this section, we only present results for one representative $N=20$ problem taken
from the 2SAT, RAN and REG class, respectively.
The plots for other $N=20$ instances look similar.

\begin{figure}[!htp]
\centering
\includegraphics[width=0.48\hsize]{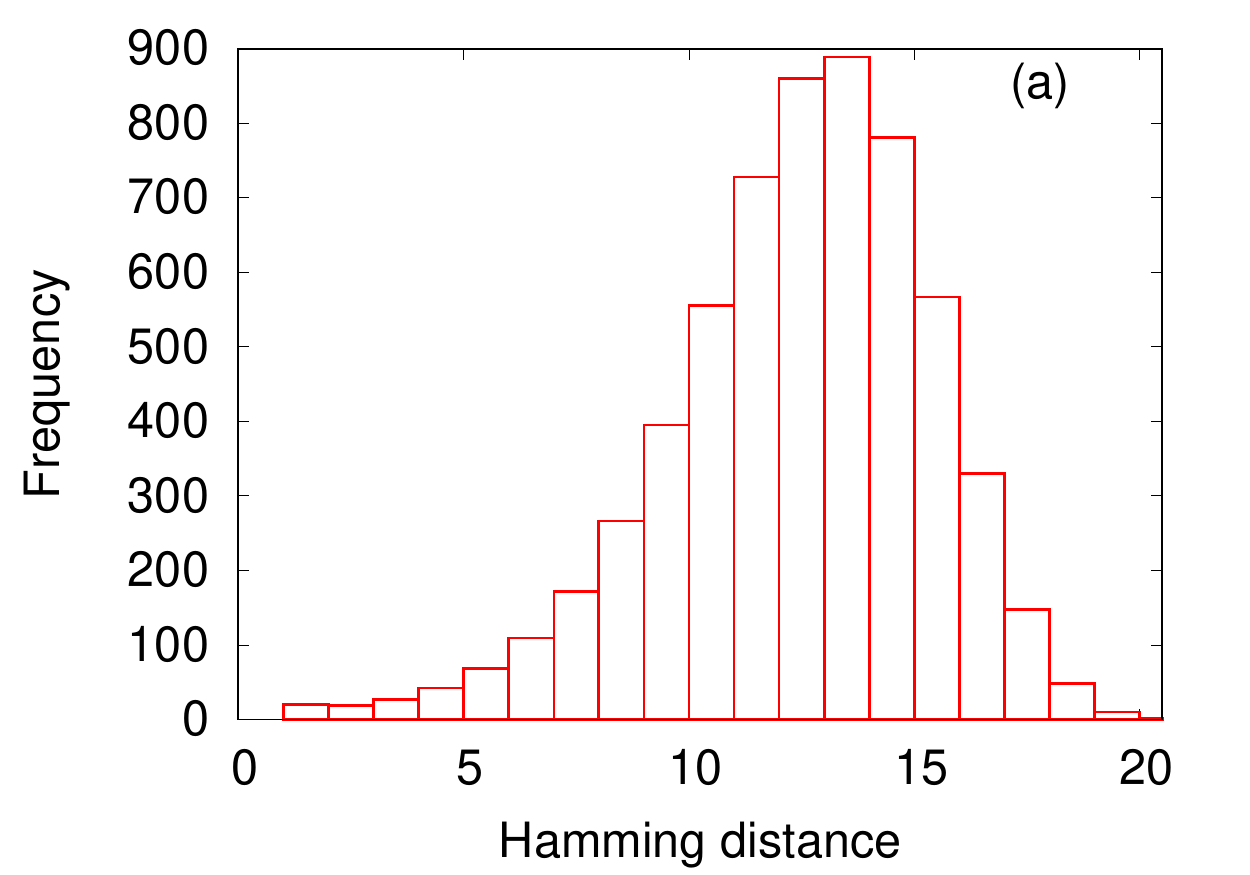}
\includegraphics[width=0.48\hsize]{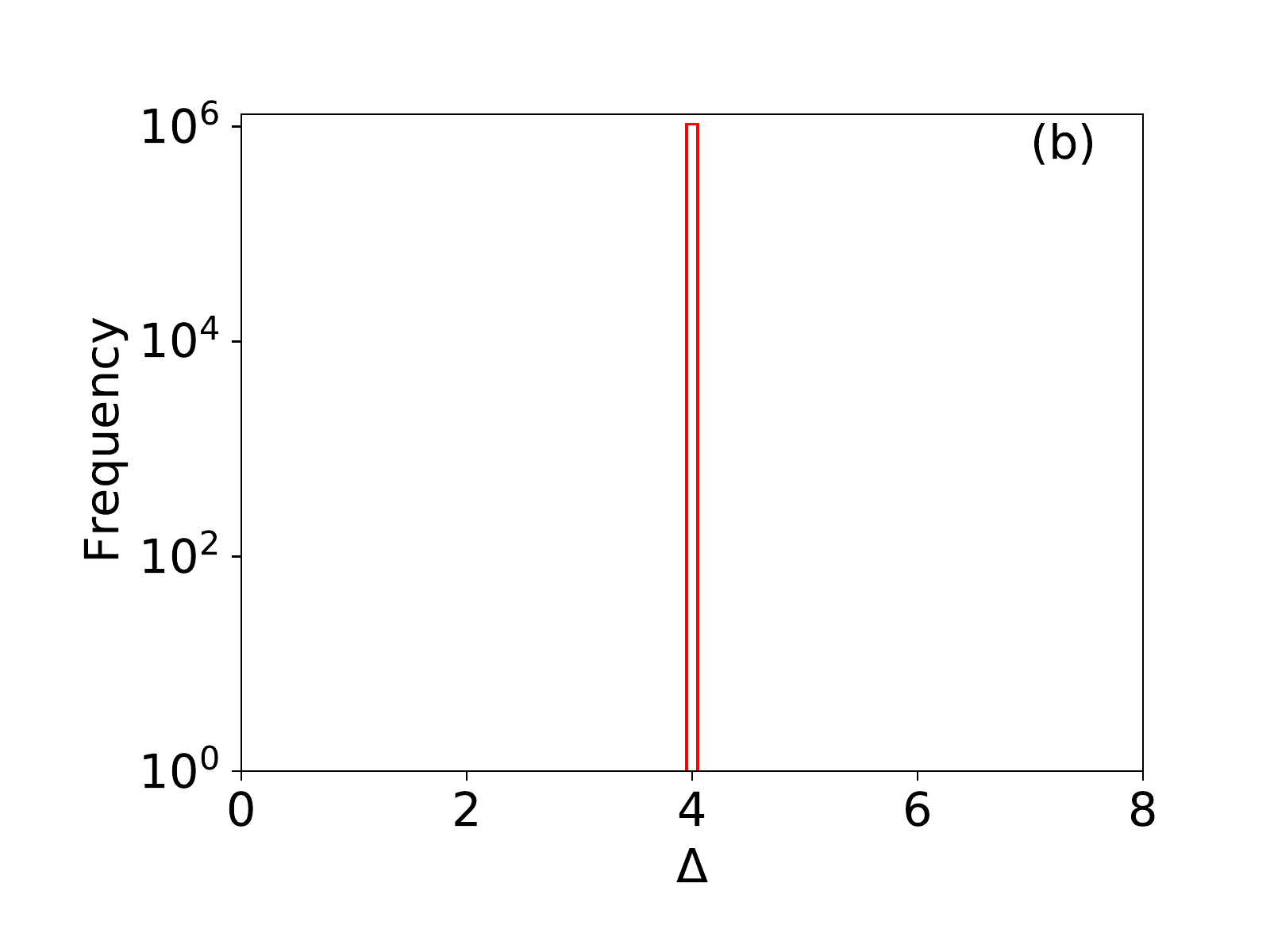}
\caption{(color online)
(a) Frequencies of Hamming distances between the ground state configuration
and the first 6036 excited states, obtained by analyzing a $N=20$ 2SAT problem.
(b) Level spacing distribution, that is the distribution
of the energy difference $\Delta$ between the first excited state and the ground state,
the second excited state and the first excited state, etc.
For the 2SAT problem considered, the first 6036 excited state are degenerate,
yielding one peak at $\Delta=4$.
\label{HAMM0}
}
\end{figure}

From Fig.~\ref{HAMM0} we draw the following conclusions.
Recall that the first excited states used to generate Fig.~\ref{HAMM0}(a) all have the same energy.
By construction, according to Eq.~(\ref{H2SAT2a}), the level spacing distribution Fig.~\ref{HAMM0}(b) is nonzero for $\Delta=4$ only.
Now imagine that the search process (in e.g., simulated annealing) for the ground state
ends up in one of these excited states.
From Fig.~\ref{HAMM0}(a) it then follows immediately that the probability
to reach the ground state by single-spin flipping will be very low.
Indeed, most of these excited states have a Hamming distance $10$-$15$
and it would require a miracle to have a particular sequence
of single-spin flips reducing the Hamming distance to zero.
In summary, from Fig.~\ref{HAMM0} it is easy to understand why
this particular class of 2SAT problems is very hard to solve by
simulated annealing~\cite{neuhaus2014monte,neuhaus2014quantum}.

\begin{figure}[!htp]
\centering
\includegraphics[width=0.48\hsize]{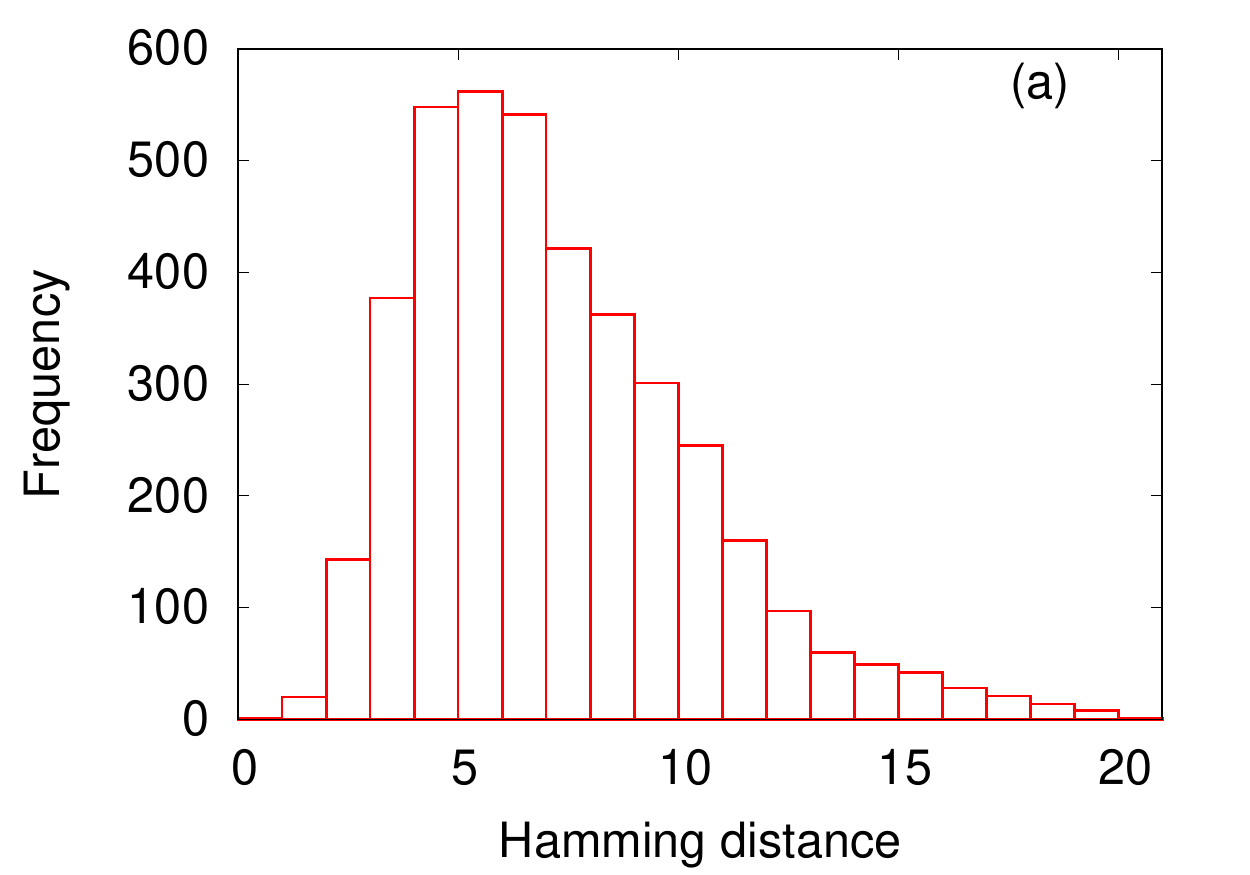}
\includegraphics[width=0.48\hsize]{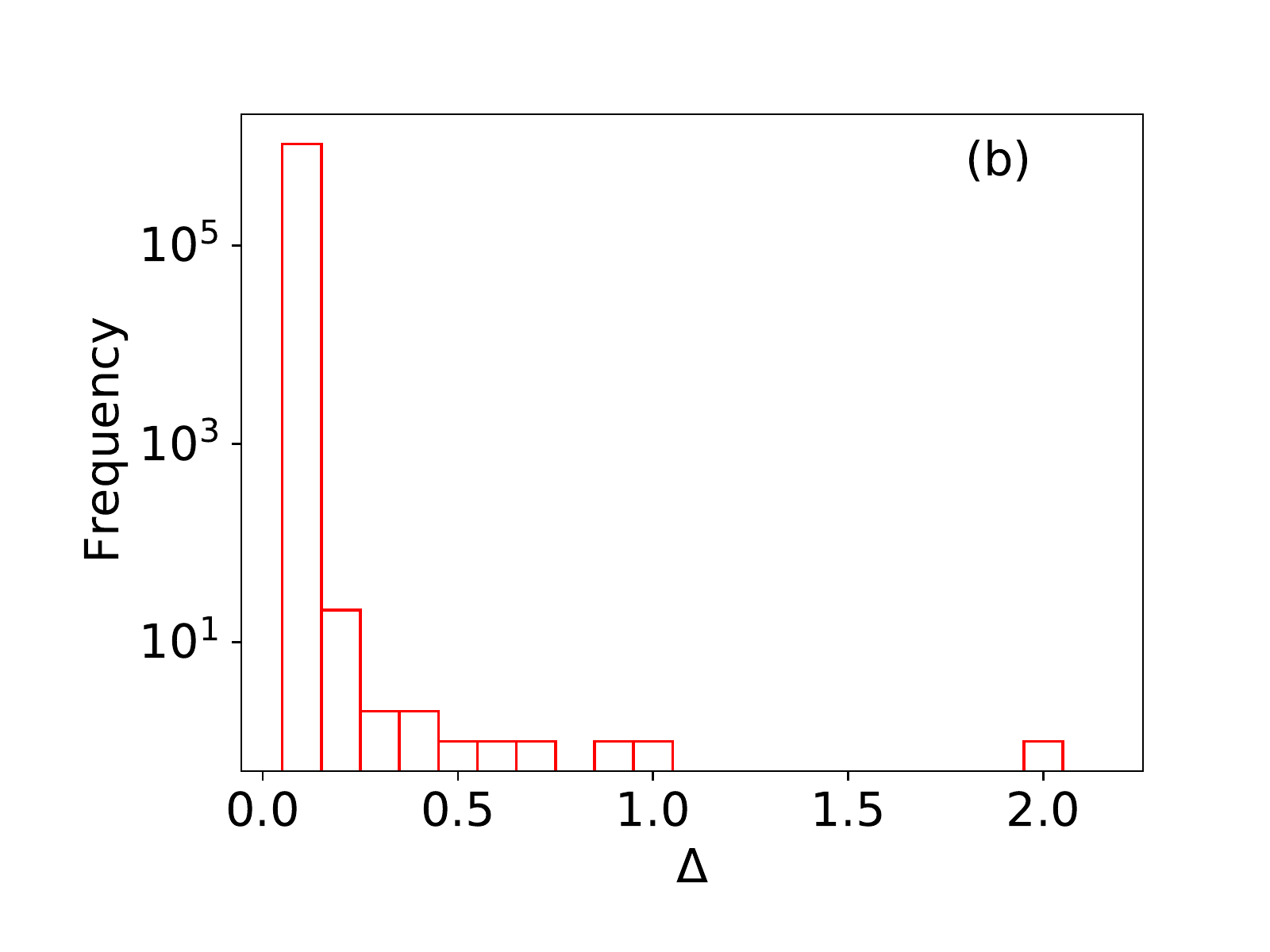}
\caption{(color online)
Same as Fig.~\ref{HAMM0} except that the problem instance belongs to the class
of $N=20$ RAN problems and that we only show the results
for the 4000 states which are closest to the ground state in energy.
In this case, the first 4000 excited energy levels differ
by approximately 17 units whereas for 2SAT, the energy  of 6036 of the lowest excited states are only 4 units of energy higher than the ground state.
\label{HAMM1}
}
\end{figure}

From Fig.~\ref{HAMM1} we conclude the following.
In contrast to Fig.~\ref{HAMM0}(a), most of the weight of the Hamming distance distribution is centered around five.
Also the level spacing distribution Fig.~\ref{HAMM1}(b) is very different from that of the 2SAT problems.
This suggests that five or less spin flips may suffice
to change the excited state into the ground state.
Thus, in comparison with the class of 2SAT problems that we have selected,
the RAN problems are expected to be much more amenable to simulated
and quantum annealing.

\begin{figure}[!htp]
\centering
\includegraphics[width=0.48\hsize]{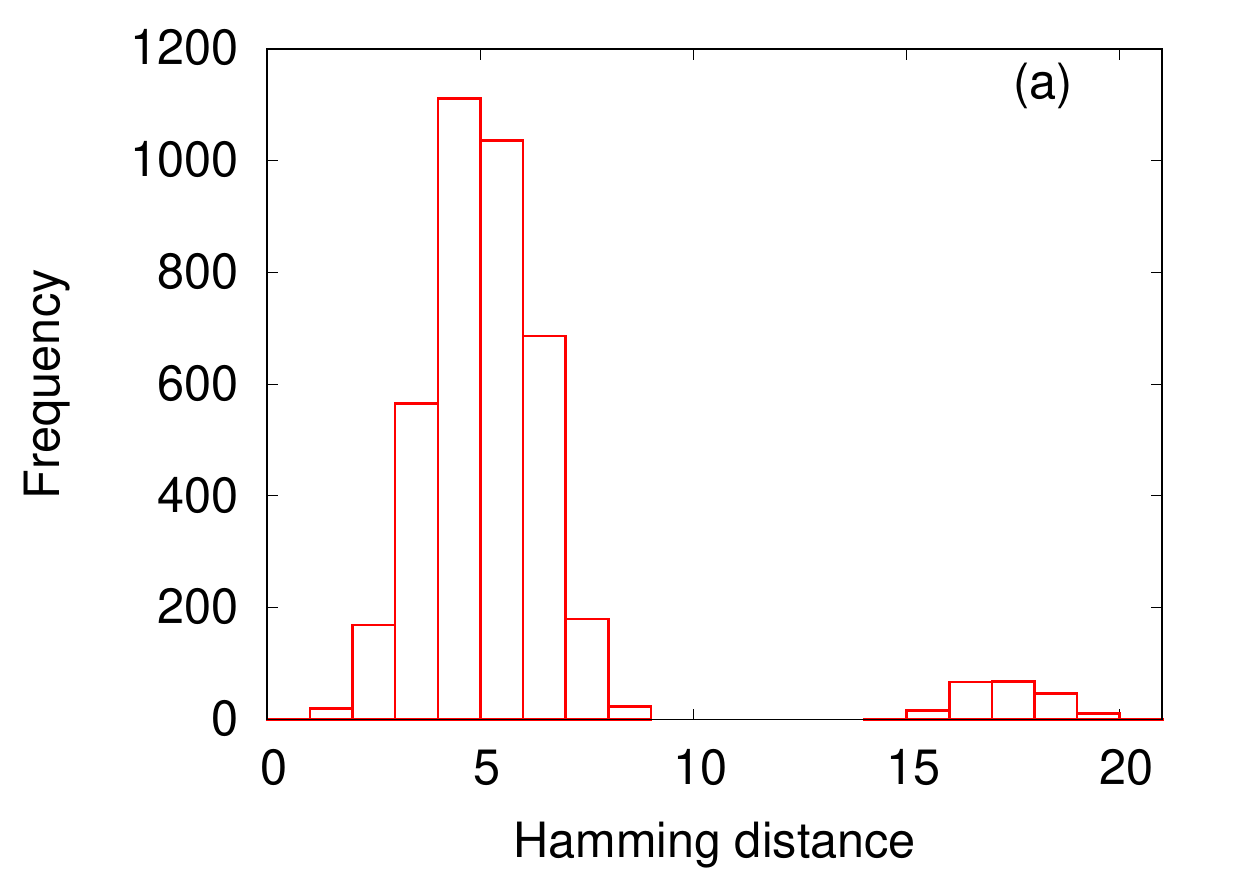}
\includegraphics[width=0.48\hsize]{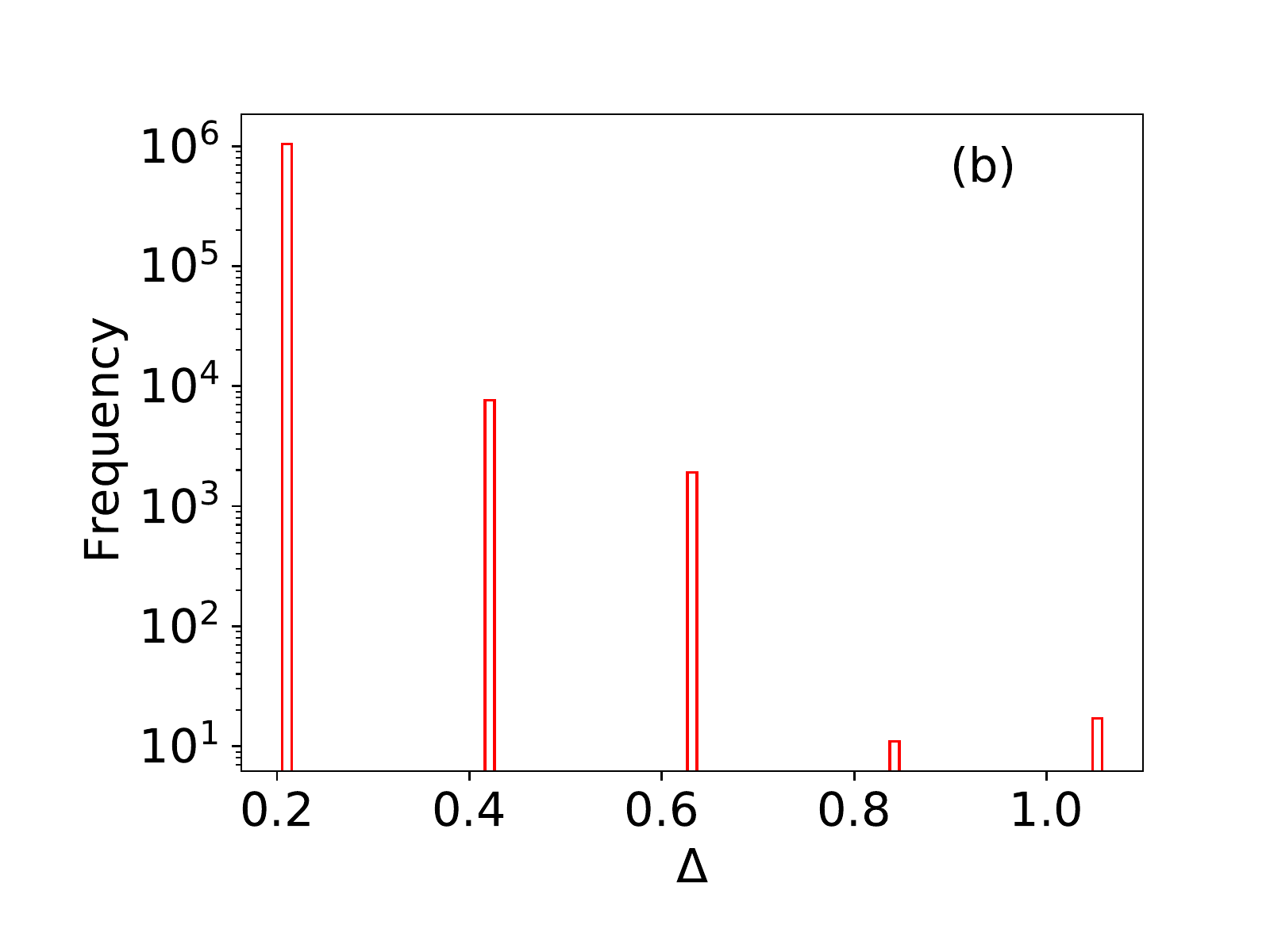}
\caption{(color online)
Same as Fig.~\ref{HAMM0} except that the problem instance belongs to the class
of $N=20$ REG problems. There are only five distinct energy differences in this case.
The energies of the 4000 lowest energy levels differ
by approximately 24 units.
\label{HAMM2}
}
\end{figure}

The Hamming distance and level spacing distribution of a REG problem (see Fig.~\ref{HAMM2}) are
very different from the corresponding ones of the 2SAT or RAN problems.
There are only five distinct energy level spacings for these problems
and the Hamming distance distribution shows that many of the excited states
differ from the ground state by only a few spin flips.
Therefore, we may expect that of the three classes of problems
considered, problems of the REG class are the least difficult to solve
by simulated or quantum annealing.

\section{Quantum annealing experiments}\label{RQA}

We demonstrate that the conclusions of section~\ref{RESU}, drawn
from the analysis of small problem instances, correlate very well
with the degree of difficulty observed when solving significantly
larger problems on a D-Wave Advantage 5.1 quantum annealer.

\begin{figure}[!htp]
\centering
\includegraphics[width=0.49\hsize]{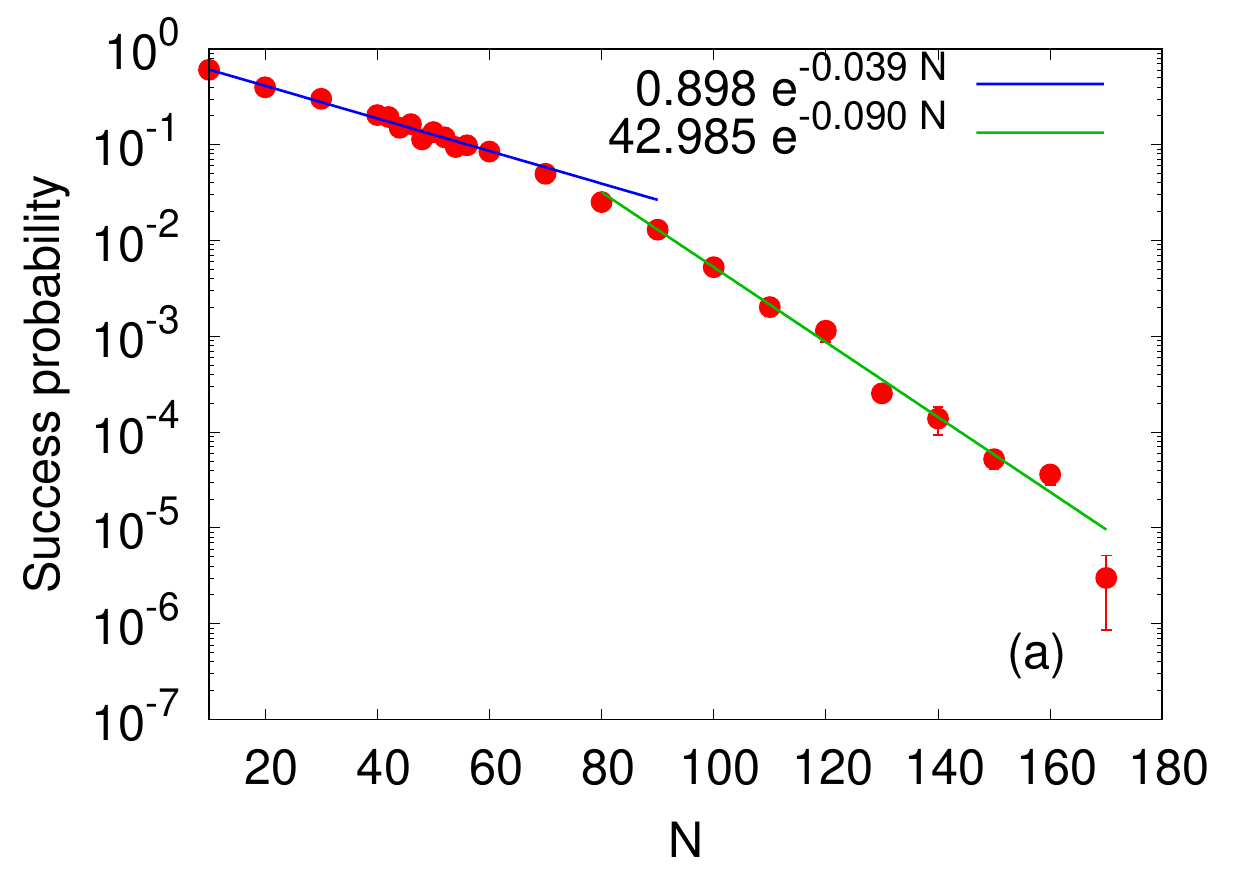}
\includegraphics[width=0.48\hsize]{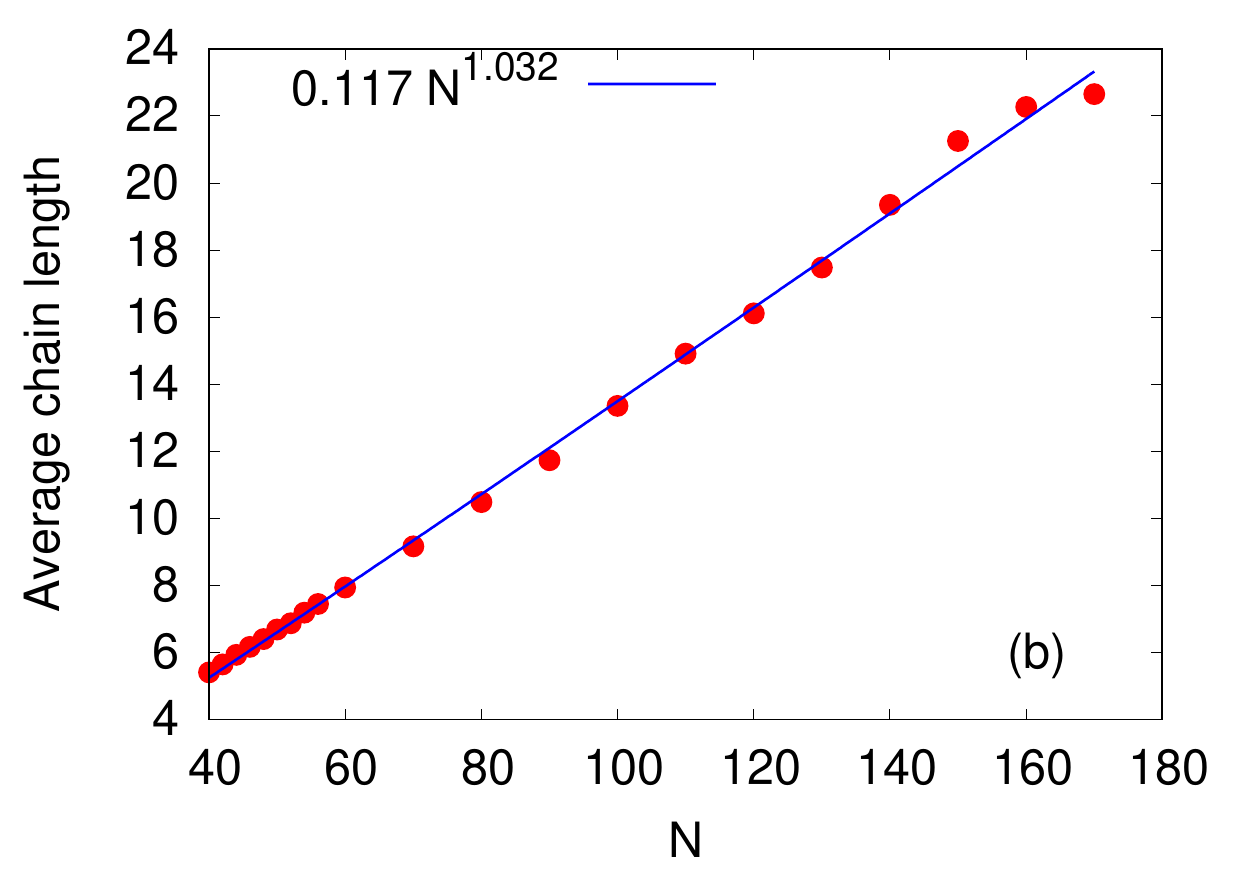}
\caption{(color online)
(a) Mean success probability and its variance as a function of the problem size $N$,
obtained by solving all problem instances of the REG class on a D-Wave Advantage 5.1 quantum annealer.
Solid lines are least square fits to data for $N<80$ and $N>80$, respectively.
(b) Average chain length as a function of the problem size $N$, a measure
for the average number of physical qubits that is required to represent one
variable in the QUBO problem.
\label{RQA1}
}
\end{figure}

In Fig.~\ref{RQA1}(a) we present the results obtained by solving all REG-class problems on a D-Wave quantum annealer.
For each $N$, instances were generated by spin-reversal transformations, as explained above.
As the problem size $N$ increases, the success probability,
that is the relative frequency with which the D-Wave yields the ground state,
decreases exponentially, from ${\cal O}(1)$ to ${\cal O}(10^{-6})$.
Fitting exponentials to the data reveals that the exponent changes from
$-0.039$ to $-0.090$ at about $N=80$.
The larger the absolute value of the exponent, the more difficult it is to solve
the QUBO by quantum annealing.

With increasing problem size $N$, it becomes more difficult and eventually impossible
to map the fully-connected QUBO problem onto the Chimera or Pegasus lattice that
defines the connectivity of the D-Wave qubits.
Even a small fully-connected problems cannot be embedded on
the D-Wave Advantage 5.1 Pegasus lattice without replacing logical bits by chains of physical qubits.
For instance, a $N=170$ variable REG problem maps onto about 3964 physical qubits of a D-Wave Advantage 5.1.
We quantify this aspect by computing the average chain length, a measure for the average number
of physical qubits that the D-Wave software uses to map a variable onto a group of physical qubits.
Figure~\ref{RQA1}(b) shows the average chain length, computed from data
obtained by solving all REG problems.
Clearly, the average chain length only increases linearly with $N$ and does not show
any sign of the crossover observed in the scaling dependence of the success probabilities, see Fig.~\ref{RQA1}(a).

\begin{figure}[!htp]
\centering
\includegraphics[width=0.48\hsize]{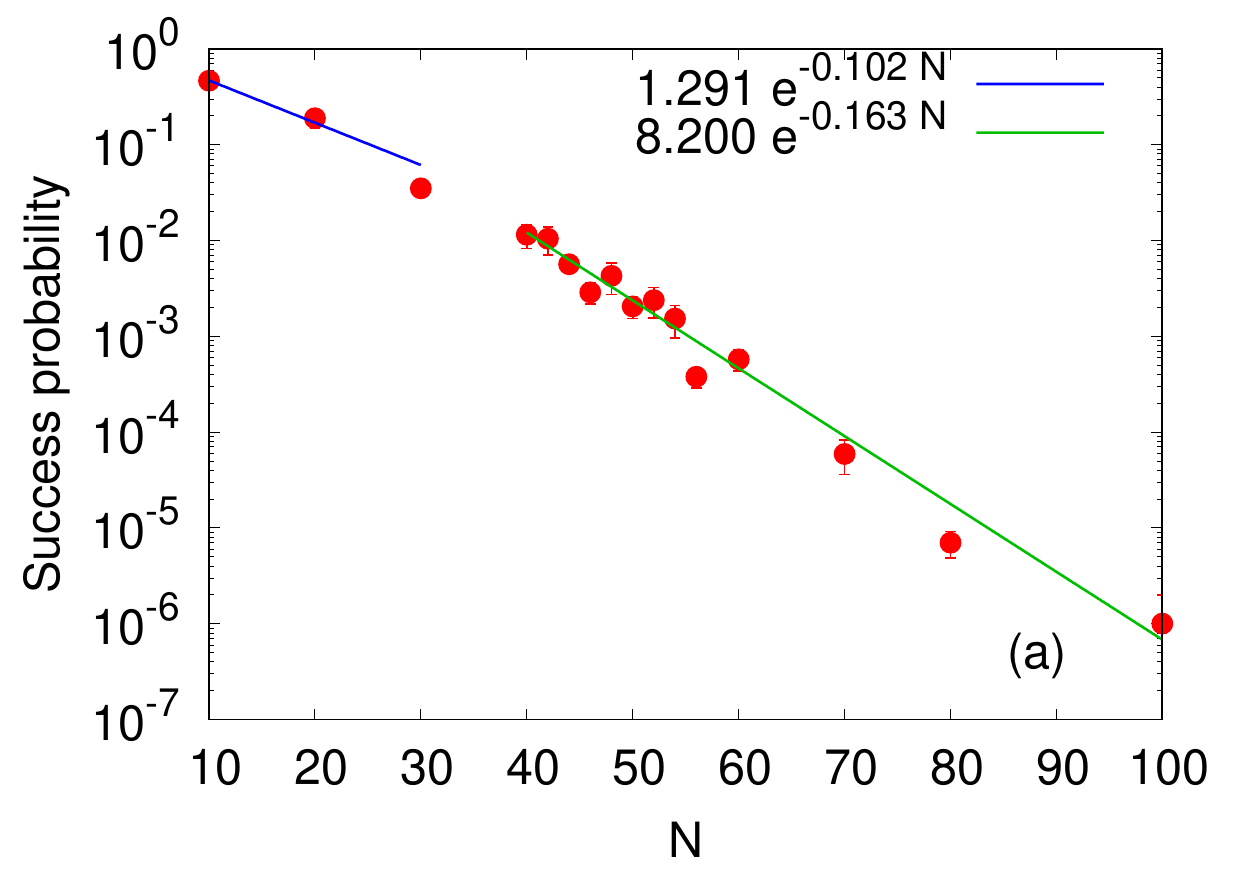}
\includegraphics[width=0.48\hsize]{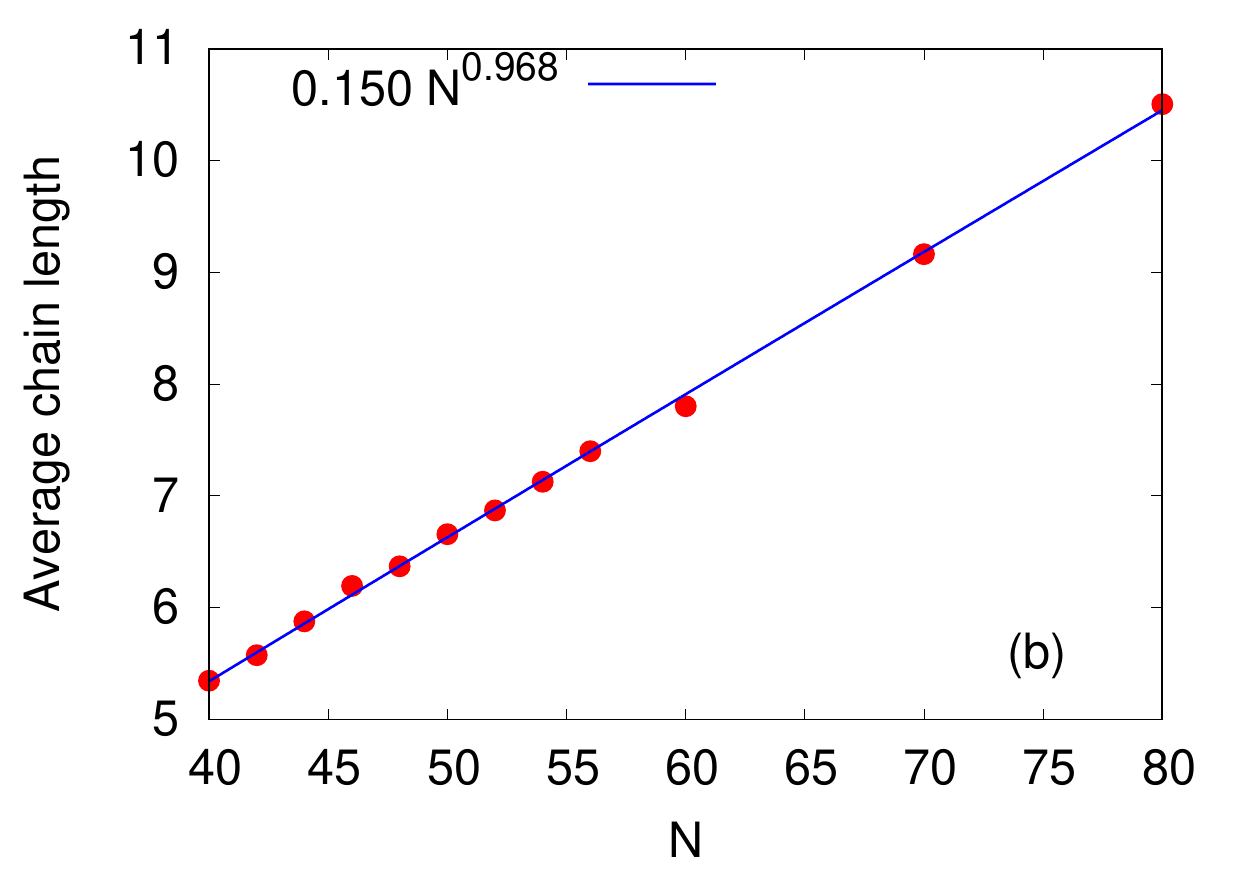}
\caption{(color online)
Same as Fig.~\ref{RQA2} except that the problem instances belongs to the RAN class.
\label{RQA2}
}
\end{figure}

In Fig.~\ref{RQA2} we present the results obtained by solving all RAN-class problems
on a D-Wave quantum annealer.
At first glance, Figs.~\ref{RQA1} and~\ref{RQA2} may look similar but there are significant differences.
First note that there is no data point for $N=90$ simply because after 200000 attempts,
the D-Wave did not return the ground state.
For $N=100$, we were more lucky but for $N\ge110$ we were not, in sharp contrast with the case
of REG problems for which the D-Wave Advantage 5.1 returned the correct solutions up to $N=170$.
Clearly, the D-Wave Advantage 5.1 quantum annealer finds RAN problems more difficult to solve than REG problems,
although both their QUBOs involve fully connected graphs.
This observation is further confirmed by fitting exponentials to the data.
As in the REG case, there is a crossover, not at $N=80$ but at $N\approx35$,
with the exponent changing from $-0.102$ to $-0.162$, see Fig.~\ref{RQA2}(a).

\begin{figure}[!htp]
\centering
\includegraphics[width=0.9\hsize]{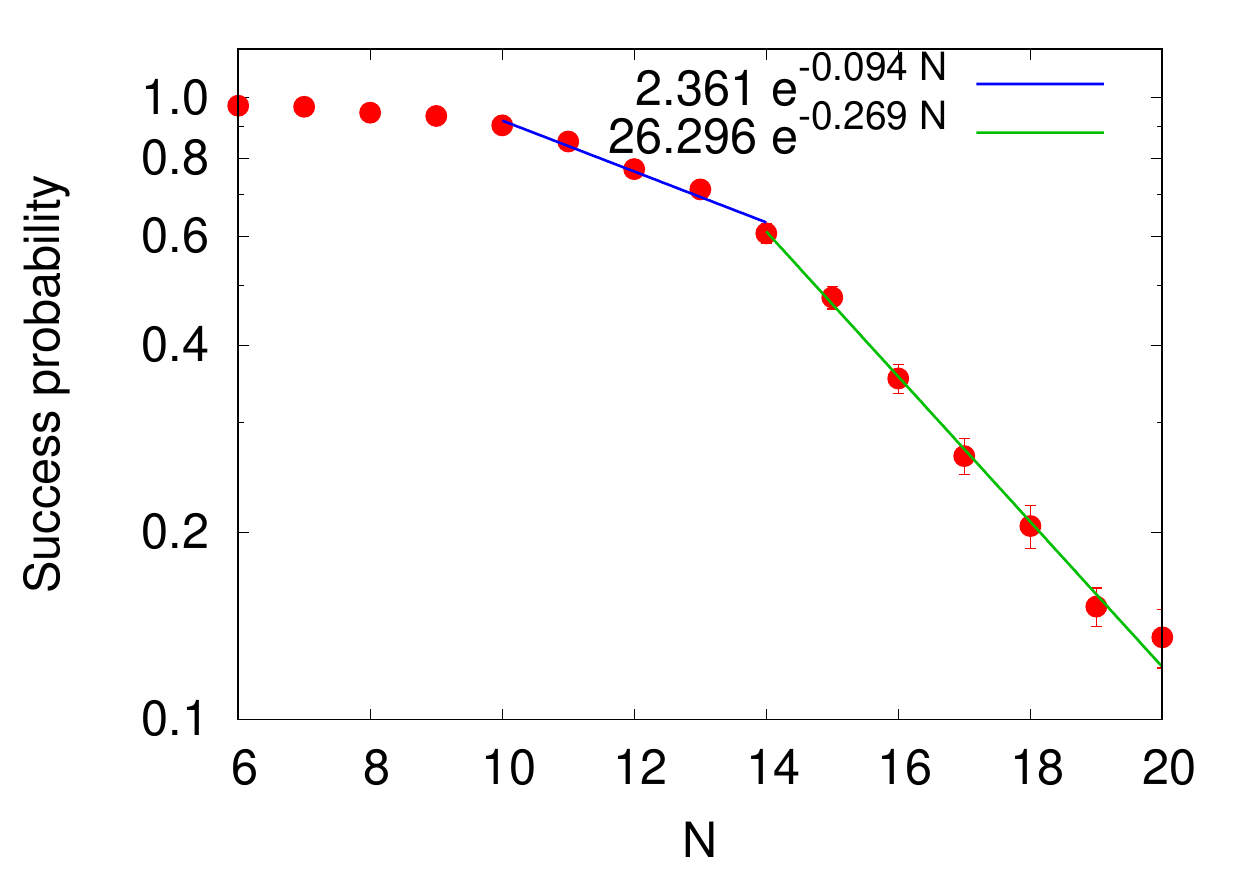}
\caption{(color online)
Same as Fig.~\ref{RQA1}(a) except that the problem instances belongs to the 2SAT class.
\label{RQA3}
}
\end{figure}

The class of 2SAT problems analyzed in this paper has already been studied extensively
through computer simulated quantum annealing and by means of D-Wave quantum annealers~\cite{Mehta2021,Mehta2022}.
The general conclusion is that, in spite of the small number ($N\le20$) of variables,
the 2SAT class contains instances which are very difficult to
solve by simulated annealing~\cite{neuhaus2014monte} and quantum annealing~\cite{Mehta2021,Mehta2022}.
For completeness, in Fig.~\ref{RQA3} we present data for the success probabilities for
the set of $N\le20$ problems, also used to study the Hamming distance and level spacing distributions.
Focusing on problem of size $N=14,\ldots,20$, we find that the success probabilities,
obtained by the D-Wave Advantage 5.1 quantum annealer, decrease exponentially with an exponent
of $-0.269$, considerable larger in absolute value than the exponents
$-0.162$ and $-0.090$ of the corresponding RAN and REG problems, respectively.
Of course, we cannot simply extrapolate the $14\le N\le20$ exponent to larger $N$'s
but, we believe it is unlikely that this exponent will become smaller as $N$ increases further.

For our D-Wave experiments, we used the default annealing time of $20 \mu\mathrm{s}$. One may expect that by using longer annealing times, the success probabilities will increase and more and larger RAN class problems can be solved.

\section{Conclusion}\label{CONC}

We have analyzed a large number of QUBOs that we have synthesized in three different ways.
The first set of QUBOs was obtained by mapping a special selection of 2SAT problems onto QUBOs.
These 2SAT problems are special in the sense that they possess a unique ground state
and a large number of degenerate, first excited states.
Findings such 2SAT problems is computationally demanding, limiting our search to instances
with less than 21 variables.

The second set of problems contains Ising models in which uniform random numbers
determine all the two-spin interactions and all the local fields.
From a statistical mechanics point of view, such fully-connected spin-glass models exhibit frustration
and computing their ground states and temperature dependent properties is known to be difficult.

Finally, the third set of problems is also of the fully-connected spin-glass type but the interaction
and field values are given by a peculiar linear function of the spin indices.
Our numerical experiments suggest that this problem may be solvable for any number of spins
but we have not yet been able to proof this conjecture mathematically.

We have calculated Hamming distance distributions and level spacing
distributions for small problem instances
and also submitted the small and large QUBO instances to a D-Wave quantum annealer.
Our results demonstrate that the exponents
characterizing the success probability of a D-Wave annealer to solve a QUBO
correlate very well with the predictions based on the Hamming distance and level spacing distributions
computed for small QUBO instances.

It would be of interest to see if Hamming distance distributions for small problem instances
also predict the effectiveness of simulated annealing.

\section*{Acknowledgements}
The authors gratefully acknowledge
support from the project JUNIQ which has received funding from the German Federal Ministry of Education
and Research (BMBF) and the Ministry of Culture and Science of the State of North Rhine-Westphalia and from
the Gauss Centre for Supercomputing e.V.
(www.gauss-centre.eu) for funding this project by providing computing time on
the GCS Supercomputer JUWELS at J\"ulich Supercomputing Centre (JSC).

\appendix

\section{Relation between QUBOs and Ising models}\label{appB}
The relations between the $Q_{i,j}$'s, $J_{i,j}$'s, $h_{i}$'s and $C_0$ are given by
\begin{eqnarray}
J_{i,j}&=&\frac{1}{4}Q_{i,j} \quad\hbox{if}\quad i\not=j\quad,\quad J_{i,i}=0 % checked 04-Feb-22
\;,
\\
h_{i}&=&-\frac{1}{2}\left(Q_{i,i}+\frac{1}{2}\sum_{j\not=i}Q_{i,j}\right)  % checked 04-Feb-22
\;,
\\
C_0&=&\frac{1}{2}\sum_{i=1}^N Q_{i,i}+\frac{1}{4}\sum_{1\le i<j\le N}Q_{i,j}
\nonumber \\
&=&\sum_{1\le i<j\le N}J_{i,j} - \sum_{i=1}^N \left[h_i+\sum_{j\not=i}J_{i,j}\right] % checked 04-Feb-22
\;,
\\
Q_{i,j}&=&4J_{i,j}-2\delta_{i,j}\left(h_i+\sum_{k\not=i} J_{i,k}\right)
\;.
\label{QUBO2}
\end{eqnarray}

\section{Exact QUBO solver}\label{appA}

We briefly discuss the scaling behavior the QUBO solver {\bf QUBO22} with the problem size $N$
and also give an impression of the resources that are required to solve QUBO problems exactly.
Not only does {\bf QUBO22} gives us the true ground state but it has also been found
very useful to identify processing units that do not perform according to specifications.
This is because the algorithm can make very close to 100\%,
sustained use of all available GPUs or CPUs, putting some severe strain on e.g., the cooling system.

The number of arithmetic operations required to solve a QUBO is proportional to $N(N-1)2^N$.
The key to ``fast'' solution is to distribute the work over many processing units.
For full enumeration, this is close to trivial.
We only have to distribute disjoint, approximately equally large, subsets of the set $\{0,\ldots,2^N-1\}$
over the available number $M$ of independent processes.
Each process enumerates its own subset to find the configurations
that corresponds to the lowest, next to lowest and largest cost.
This step takes $\hbox{OpCount}(N/M)$ operations per process.
The results of all processes are then gathered and
used to find the lowest, next to lowest and largest cost of the full set.
The latter step takes ${\cal O}(M)$ operations.

Solving a $N=40$ QUBO using 32 Intel Xeon Platinum 8168 CPUs with $24$ cores each takes about 505 s.
Solving the same QUBO using 4 NVidia A100 GPUs take about 82 s.
Roughly speaking, for solving QUBOs, 1 NVidia A100 GPUs has the computational power of 49 Intel Xeon Platinum 8168 CPUs.
Obviously, using GPUs instead of CPUs reduces the elapsed time to solve QUBO problems significantly.

In Table~\ref{TAB1} we present strong- and weak-scaling results for {\bf QUBO22}
running on the GPUs of JUWELS booster~\cite{BOOSTER}.
The problem instances are fully connected, regular QUBOs with $N$ variables.
Note that the measured elapsed times for $44<N\le54$ are a little {\bf smaller} than
the corresponding predictions based on the $N=44$ elapsed time,
demonstrating the close-to-ideal strong- and weak-scaling behavior of {\bf QUBO22}.

\begin{table}[!htp]
\caption{
Strong- and weak-scaling results for {\bf QUBO22} using NVidia A100 GPUs.
The second column gives the number of GPUs used and the third column lists the elapsed time to solution.
The ratio of arithmetic operation counts defined by $S(44)=N(N-1)2^N/(44\times43\times2^{44})$ (last column),
the ratio of the number of GPUs used,
and the elapsed time for solving the $N=44$ QUBO are combined to predict the elapsed time (fourth column) for $N>44$.
}
\begin{center}
\begin{ruledtabular}
%\begin{tabular}{rrrrrrrrrrrr}
\begin{tabular}{lrrrr}
$N$      & A100 GPUs & Elapsed time (s)& $1562 \times S_{44}(N)$ (s) & $S_{44}(N)$ \\%
\hline\noalign{\vskip 4pt}
 44    & 4    & 1562  & 1562  &  1    \\
 50    & 256  & 1973  & 2015  &  1.29 \\
 50    & 512  &  987  & 1015  &  0.65 \\
 50    & 1024 &  493  &  500  &  0.32 \\
 54    & 1024 & 9121  & 9451  &  6.05 \\
 56    & 512 &  83502 &81224  &  52 \\
\end{tabular}
\end{ruledtabular}
\label{TAB1}
\end{center}
%\end{table*}
\end{table}
%\end{widetext}

\bibliography{qub0}
\end{document}